UCM-DFT-11/94

# A Proposal of a Renormalization Group Transformation for Lattice Field Theories

L. A. Fernández, A. Muñoz Sudupe, and J.J. Ruiz-Lorenzo*,

*Departamento de Física Teórica, Universidad Complutense de Madrid, 28040 Madrid, Spain,*

(e-mail: laf, sudupe, ruiz@lattice.fis.ucm.es)

A. Tarancón,

*Departamento de Física Teórica, Universidad de Zaragoza, 50009 Zaragoza, Spain.*

(e-mail: tarancon@sol.cie.unizar.es)

## Abstract

We propose a new Real Space Renormalization Group transformation useful for Monte Carlo calculations in theories with global or local symmetries. From relaxation arguments we define the block-spin transformation with two tunable free parameters, adapted to the system's action. Varying them it is possible to place the fixed point very close to the simulation point.

We show how the method works in a simple model with global symmetry: the three dimensional XY model.

*and Dipartimento di Fisica, Università di Roma I, P. A. Moro 2, 00185 Roma, Italy

## I. INTRODUCTION

Real Space Renormalization Group (RSRG) methods have become an extremely useful tool for understanding critical phenomena. The use of the Renormalization Group (RG) ideas in the framework of Monte Carlo simulation has been very successful. However there are some difficulties that restrict their use, specially in gauge theories.

The main problem is the necessity of using many couplings to describe the RG trajectory after several scale transformations. In the case of gauge theories, the preservation of the local symmetry adds a further difficulty in the definition of the Renormalization Group Transformation (RGT).

To avoid the appearance of many new significant couplings, we need to improve the RGT in order to get the RG fixed point closer to the simulation point. In this way the generated couplings are of relatively less importance and thus the truncation errors are strongly reduced [1]. This idea has been applied to spin [1] and gauge [2] theories. In these works a sum over neighbor spins or over different paths is made. The mean is weighted depending on some free parameters which can be tuned.

Consequently, the way of constructing efficiently the Renormalized Fields (RF) is a key problem. While the mean over neighbors gives good results in simple models, it becomes more involved with complex actions.

The situation gets worse when considering gauge theories. In such a case the necessity of preserving gauge invariance forces to take the mean over ordered products of fields along fixed–ends trajectories. This calculation is in practice carried out only for close end–points, because, otherwise, the number of needed trajectories becomes very large. When computing in parallel machines this procedure can become very time consuming. Moreover the chosen trajectories should not leave the considered block, in order to avoid the exchange of information among them.

On the other hand it is well known the existence of powerful relaxation techniques in the study of several problems like spectroscopy [3] or topological studies [4]. From the point of view of spectroscopy calculations, the idea is to reduce the short distance fluctuations, thus obtaining a better projection of the operators over the desired physical state.

According to the actual form of the action an *ad hoc* transformation of the fields (smearing) is built that damps the high frequencies. We remark that in the gauge case it has sense only to consider the smoothing of the energy distribution or of any other gauge invariant operator, not of the fields themselves, as the local symmetry makes meaningless the concept of local value of the field.

Our proposal for a RGT consists in a two step procedure. We first perform a relaxation transformation, suitable both for spin and gauge theories, and then, a simple change of scale (blocking) by a factor of two in order to define the RF. There are free parameters in the transformation that allow us to place the fixed point of the RGT over a wide region of the coupling space. The best choice corresponds to place it close to the simulation point. By iterating the RGT it is possible to reach lattice sizes as small as desired; down to side $L = 2$ if we start, as will be done in this article, from lattices with side $L = 2^l$.

The study of the coupling flux in the parameter space is useful to compute the fixed points and critical exponents [5]. Using the Schwinger–Dyson Equations (SDE) on the lattice [6] it is also possible to measure the renormalized coupling at every RG step.

In this paper we use the 3–dimensional XY model, with well known phase diagram and with critical exponents accurately measured, to simplify the discussion and adjust the method. We will particularize the notation to this case. We remark that the main usefulness of this proposal lies in the framework of a gauge theory. Our attention will be focused in what can be learned in this simple model, namely: how to choose the better transformation, how to estimate the systematic errors, which is the best way to reduce the statistical errors, etc..

In section 2 we present the details of the method, leaving the discussion of the Schwinger–Dyson equations for section 3. In section 4 we study the flux diagram and fixed point location, computing the critical exponent $\nu$ in section 5. A Finite Size Scaling analysis is shown in section 6. Finally section 7 is devoted to conclusions.

## II. RGT: SMEARING AND BLOCKING

### A. Smearing

Let us consider a real scalar field $\varphi(\boldsymbol{x},\tau)$ where $\boldsymbol{x}$ belongs to a $d$-dimensional space. One method [3] to damp the high spatial frequencies of a given configuration is to consider the evolution driven by the heat equation

$$\frac{\partial\varphi(\boldsymbol{x},\tau)}{\partial\tau} = \kappa\Delta\varphi(\boldsymbol{x},\tau), \tag{1}$$

whose solution in terms of the Fourier transform $\widehat{\varphi}(\boldsymbol{k},\tau)$ is

$$\widehat{\varphi}(\boldsymbol{k},\tau) = \widehat{\varphi}(\boldsymbol{k},0)e^{-\kappa\tau\boldsymbol{k}^2}. \tag{2}$$

In this way, with an appropriate selection of $\kappa$ and $\tau$ it is possible to eliminate frequencies higher than a desired cutoff.

In practice to compute a $\tau$-evolution following equation (1) is very easy in the lattice. Calling $\varphi_{\boldsymbol{n},s} \equiv \varphi(\boldsymbol{n}a, sb)$, after a discretization of the Laplace operator we obtain the following iterative scheme

$$\varphi_{\boldsymbol{n},s+1} = \varphi_{\boldsymbol{n},s} + \epsilon\sum_{\mu}(\varphi_{\boldsymbol{n}+\boldsymbol{\mu},s} + \varphi_{\boldsymbol{n}-\boldsymbol{\mu},s} - 2\varphi_{\boldsymbol{n},s}), \tag{3}$$

where $\epsilon = \kappa b/a^2$, $\mu = 0,\ldots,d-1$ and $\boldsymbol{\mu}$ is the unit vector in the $\mu$ direction.

For a general system those equations are substituted by any iteration that locally reduces the energy. This process depends on the form of the action and is not univocally determined; moreover, the variables may belong to a compact group and, in order to keep them inside it, we may have to project them back in a specific way. Another possibility, that avoids the projection over the group, is to work with variables outside it, in this case however the *a priori* unknown anomalous dimension of the new fields should be considered in order to find the fixed point.

In many spin systems as well as in gauge theories, we can schematically write the partition function as

$$Z = \int [dg] \exp\left\{ \beta\, \chi \left( \sum_{\{\rho\}} g_\rho \sum_{\{\sigma\}} h^\dagger_{\rho\sigma} \right) \right\}, \tag{4}$$

where $g_\rho$ and $h_{\rho\sigma}$ belong to a compact Lie group, and $\chi$ is a character function of the considered group. In spin models $h_{\rho\sigma} = g_{\rho+\sigma}$, and $\{\sigma\}$ extends over the nearest neighbors in the forward direction for instance. On gauge models $\{\rho \equiv (\boldsymbol{n}, \mu)\}$ stands for all links, $h_{\rho\sigma} = g^{staple}_{\rho\sigma}$, and $\{\sigma\}$ extends over the staples connected to the link $\rho$.

Although the proposed methods are general for spin or gauge and abelian or non abelian systems, in order to simplify the notation hereafter we will restrict ourselves to U(1), where $g = e^{i\theta}$ with $\theta \in (-\pi, \pi]$. By the same reason we will consider the fundamental representation, *i.e.* the more simple action, with $\chi(g) = \mathrm{Re}\; g = \cos\theta$.

The simplest generalization of equation (3) is

$$g_{\rho,s+1} = \mathcal{P}\left[ g_{\rho,s} + \epsilon \sum_{\{\sigma\}} h_{\rho\sigma,s} \right], \tag{5}$$

where $\mathcal{P}$ means the projection over the group (division by the modulus in the present case).

This transformation is performed in all lattice sites in such a way that in the computation of $g_{\rho,s+1}$ in (5) only the variables at smearing step $s$ are used, even though some neighbor sites could have been already modified. The variation of the energy computed changing $g_{\rho,s} \to g_{\rho,s+1}$ without modifying $h_{\rho\sigma,s}$ is always negative. However, after a whole sweep, when all variables are changed, the reduction is expected only for the mean value of the energy.

In disordered configurations $g_\rho h^\dagger_{\rho\sigma}$ is not near to 1 (in the XY model $\left\langle g_\rho h^\dagger_{\rho\sigma} \right\rangle \approx 0.3$ near the critical point), and there are not clear *a priori* arguments for selecting (5) between many other transformations.

In fact we will use the following one

$$g_{\rho,s+1} = g_{\rho,s} \left[ \mathcal{P}(g^\dagger_{\rho,s} \sum_{\{\sigma\}} h_{\rho\sigma,s})^\epsilon \right], \tag{6}$$

where for the $\epsilon$-power definition we select the argument of the basis in the $(-\pi, \pi]$ interval (for other groups we would select suitable symmetric regions). If we write $\sum_{\{\sigma\}} h_{\rho\sigma} = C e^{i\Theta_\rho}$, it is easy to see that the local reduction of the energy, which always holds, does not depend on the factor $C$, thus its smoothing intensity is similar for disordered (small $C$) or ordered configurations.

We have numerically found that the transformation (6) performs better than (5) regarding the stability of the observables. All the numerical results presented in this paper have been obtained with the transformation (6). We will present in section IV some numerical results about the performance of the procedure as a function of $\epsilon$ and the number of relaxation iterations $n_s$.

### B. Blocking

The relaxation procedure considered above does only half of the work needed in a RGT. After it, the high frequencies have been damped out and relevant (low frequency) information has been propagated along the lattice.

After the relaxation procedure all renormalized fields at small distances are nearly equal, as we have fluctuations only at large distances (or small momenta). This makes nearly irrelevant the sum over paths or over different points for spin systems, and therefore we can follow a simple decimation procedure to perform the blocking transformation without a significant loss of information. For a gauge theory the decimation consists in replacing the product $U_{2\boldsymbol{n},\mu} \cdot U_{2\boldsymbol{n}+\boldsymbol{\mu},\mu}$ by a new link of the blocked lattice, discarding the rest.

Our complete RGT consists then of the following steps:

1. On the original lattice we perform $n_s$ iterations with (6).

2. We block the system by a factor 2, using decimation.

We are then left with two free parameters, $n_s$ and $\epsilon$, which permit us to control the position of the fixed point inside the critical surface.

Beginning from a cubic ($L^d$) lattice with $\log_2 L$ integer, after iterating the RGT up to a blocked lattice side equal to $L = 2$ we have a sequence of sizes $\{N_b = L/2^b\}$ and renormalized fields and couplings $\{\{\theta^b\},\beta^b\}$ where $b=0,\ldots,\log_2 L$ is the block level with $b=0$ being the original lattice.

On the original lattice the dynamics is governed by the value of the unrenormalized $\beta_i$ parameters. In the blocked lattices the distribution of the fields comes from the original distribution and from our RGT.

We can compute on these lattices not only the observables but also the couplings needed in order to obtain the same values for the observables in an independent simulation: the renormalized couplings (see next section).

Starting from a lattice of side $L$ with couplings $\boldsymbol{\beta}=(\beta_1,\ldots,\beta_n,\ldots)$ ($N_0$ and $\boldsymbol{\beta}^0$ respectively in the previous notation) we arrive to $N_1$ and $\boldsymbol{\beta}^1$ after a RGT. The movement from $\boldsymbol{\beta}^0$ to $\boldsymbol{\beta}^1$ represents the RG flux starting from $\boldsymbol{\beta}^0$ after a RGT with a scale change of 2. This discussion applies to all levels of RGT. Once on the fixed point the system does not evolve anymore. We remark that in order to accomplish that, it is crucial that all the steps must be identical at all the blocking levels.

## III. SDE FOR THE XY MODEL

We will apply, as an example, the precedent method to the three dimensional XY model. The conclusions that we will obtain will be hopefully of a wider generality.

The partition function for that model is

$$Z=\int[d\theta]\exp\left\{\beta_1\sum_{\boldsymbol{n},\mu}\cos(\theta_{\boldsymbol{n}}-\theta_{\boldsymbol{n}+\boldsymbol{\mu}})\right\}, \tag{7}$$

where $\boldsymbol{\mu}$ is the unitary vector in the the $\mu$ direction and the sum in $\mu$ extends from 0 to $d-1$.

In $d=3$ this model has a second order phase transition, with a global symmetry breaking for $\left\langle e^{i\theta}\right\rangle$, at $\beta_{1,c}=0.45420(2)$ and thermal critical exponent $\nu$ in the range $0.66-0.67$ [7,8]. We will use these values to compare with our computation.

In general, when we perform a RGT new couplings will be generated in the system. Our goal will be that after iterating the RGT in the XY model only the nearest neighbors coupling, $\beta_1$, will be significantly non zero, after an appropriate selection of the smearing parameters $n_s$ and $\epsilon$.

In order to check it, let us suppose that this is not the case, and compute more renormalized couplings to see if they are effectively zero. We will compute only the next to nearest renormalized coupling, that is, an interaction between neighbors at a $\sqrt{2}$ distance. The calculation of further couplings is more involved because it suffers from more numerical uncertainty and we will assume that this test is sufficient for our purposes.

The partition function when the two couplings are considered is

$$Z = \int [d\theta] \exp \left\{ \beta_1 \sum_{\boldsymbol{n},\mu} \cos(\theta_{\boldsymbol{n}} - \theta_{\boldsymbol{n}+\boldsymbol{\mu}}) + \beta_2 \sum_{\boldsymbol{n},\mu<\nu} \cos(\theta_{\boldsymbol{n}} - \theta_{\boldsymbol{n}+\boldsymbol{\mu}+\boldsymbol{\nu}}) \right\}. \tag{8}$$

As the number of neighbors in $d = 3$ is twice as much for the $\beta_2$ interaction as for $\beta_1$, the phase diagram in the region where both $\beta_1$ and $\beta_2$ are positive (where there is no frustration), will consist of two phases: ordered and disordered, separated by a nearly straight line with slope $-\frac{1}{2}$ that goes trough the point $(0.45420, 0)$ (see figure 1).

Let us compute the SDE for this two couplings system following the procedure proposed in [6]. Let $A(\theta)$ be a function with null expectation value. This trivially implies that also $\partial \langle A(\theta) \rangle / \partial\theta_{\boldsymbol{n}} = 0$.

At a certain blocking level $b$ of RGT, we will have a large number of non zero couplings and $Z$ will take the form

$$Z = \int [d\theta^b] \exp\{-\sum_i \beta_i^b S_i(\theta^b)\}, \tag{9}$$

where $S_i$, function of the renormalized fields, is the action corresponding to the renormalized coupling $i$ at level $b$.

We have

$$\left\langle A(\theta^b) \right\rangle = Z^{-1} \int [d\theta^b] A(\theta^b) \exp\{-\sum_i \beta_i^b S_i(\theta^b)\} = 0, \tag{10}$$

and then we obtain the following identity

$$0 = \frac{\partial \left\langle A(\theta^b) \right\rangle}{\partial \theta^b_{\boldsymbol{n}}} = -Z^{-1} \frac{\partial Z}{\partial \theta^b_{\boldsymbol{n}}} \left\langle A(\theta^b) \right\rangle + \left\langle \frac{\partial A(\theta^b)}{\partial \theta^b_{\boldsymbol{n}}} \right\rangle$$
$$- \sum_i \beta^b_i \left\langle A(\theta^b) \frac{\partial S_i(\theta^b)}{\partial \theta^b_{\boldsymbol{n}}} \right\rangle . \tag{11}$$

Taking into account that $\left\langle A(\theta^b) \right\rangle = 0$ we find

$$\left\langle \frac{\partial A(\theta^b)}{\partial \theta^b_{\boldsymbol{n}}} \right\rangle = \sum_i \beta^b_i \left\langle A(\theta^b) \frac{\partial S_i(\theta^b)}{\partial \theta^b_{\boldsymbol{n}}} \right\rangle . \tag{12}$$

This algebraic equation relates the value of $\beta^b_i$ with expectation values at a certain blocking level $b$, and then, allows us to compute the renormalized couplings from the known expectation values. These renormalized couplings, if used in $Z$, should give us the same values for all observables at each value of $b$.

We see in (12) that we need as many independent operators as non zero couplings in order to invert this equation and compute the renormalized couplings.

In the hypothesis that also for $b > 0$ only $\beta^b_1$ is different from zero, let us consider the function

$$A(\theta^b) = \sin(\theta^b_{\boldsymbol{n}} - \theta^b_{\boldsymbol{n}+\boldsymbol{\mu}}), \tag{13}$$

which, when used in equation (12), gives

$$\beta^b_1 = \frac{\left\langle \sum_{\pm\mu} \cos(\theta^b_{\boldsymbol{n}} - \theta^b_{\boldsymbol{n}+\boldsymbol{\mu}}) \right\rangle}{\left\langle (\sum_{\pm\mu} \sin(\theta^b_{\boldsymbol{n}} - \theta^b_{\boldsymbol{n}+\boldsymbol{\mu}}))^2 \right\rangle} . \tag{14}$$

This equality is exact for $b = 0$.

Let us compute now $\beta^b_1$ and $\beta^b_2$ assuming that we have two couplings, see equation (8). From (12) we need two operators to compute the couplings. One of them will be the previously used $A$ and the other could be

$$B(\theta^b) = \sin(\theta^b_{\boldsymbol{n}} - \theta^b_{\boldsymbol{n}+\boldsymbol{\mu}+\boldsymbol{\nu}}), \tag{15}$$

with $\mu \neq \nu$, that is, an operator with the same fields combination as $S^b_2$, that we hope will be the best one coupled to $\beta^b_2$ giving the best signal-noise relation. Now the equations for obtaining the RG couplings are

$$\begin{pmatrix} \langle E_1^b \rangle \\ \langle E_2^b \rangle \end{pmatrix} = \begin{pmatrix} \langle (A_1^b)^2 \rangle & \langle A_1^b A_2^b \rangle \\ \langle A_1^b A_2^b \rangle & \langle (A_2^b)^2 \rangle \end{pmatrix} \begin{pmatrix} \beta_1^b \\ \beta_2^b \end{pmatrix} , \tag{16}$$

where

$$\begin{aligned} \langle E_1^b \rangle &= \left\langle \sum_{\pm\mu} \cos(\theta_{\boldsymbol{n}}^b - \theta_{\boldsymbol{n}+\boldsymbol{\mu}}^b) \right\rangle , \\ \langle E_2^b \rangle &= \left\langle \sum_{\pm\mu,\pm\nu,\mu<\nu} \cos(\theta_{\boldsymbol{n}}^b - \theta_{\boldsymbol{n}+\boldsymbol{\mu}+\boldsymbol{\nu}}^b) \right\rangle , \\ A_1^b &= \sum_{\pm\mu} \sin(\theta_{\boldsymbol{n}}^b - \theta_{\boldsymbol{n}+\boldsymbol{\mu}}^b) , \\ A_2^b &= \sum_{\pm\mu,\pm\nu,\mu<\nu} \sin(\theta_{\boldsymbol{n}}^b - \theta_{\boldsymbol{n}+\boldsymbol{\mu}+\boldsymbol{\nu}}^b) . \end{aligned} \tag{17}$$

From this expression, it is possible to compute $\beta_1^b, \beta_2^b$, by inverting the $2 \times 2$ matrix in (16). With these values we can draw the flux diagram of the model on the $(\beta_1, \beta_2)$ plane, which permits us to determine the fixed point in this plane for a concrete RGT prescription.

The use of these techniques allows the determination of the whole flux diagram of the model, the number of couplings being limited by the numerical precision not by the method itself.

## IV. FLUX DIAGRAM: FIXED POINT LOCATION

Let us consider a point in the parameter space $\boldsymbol{\beta}^b \equiv (\beta_1^b, \beta_2^b, \ldots)$ corresponding to the RG block level $b$. If this point is near to the fixed point $\boldsymbol{\beta}^\star \equiv (\beta_1^\star, \beta_2^\star, \ldots)$ the equations for the RG transformation $\boldsymbol{\beta}^{b+1} - \boldsymbol{\beta}^\star = T(\boldsymbol{\beta}^b - \boldsymbol{\beta}^\star)$ can be linearized.

To fix the notation let us call $\boldsymbol{e}^\alpha$ and $\lambda_\alpha = s_\alpha^y$ respectively to the eigenvectors and eigenvalues of the matrix $T$, $s$ being the change of scale. So that we can write

$$\boldsymbol{\beta}^b - \boldsymbol{\beta}^\star = \sum_\alpha t_\alpha^b \boldsymbol{e}^\alpha, \tag{18}$$

where $t_\alpha^b$ are the scaling fields at blocking level $b$ that after a RGT transform as $t_\alpha^{b+1} = s^{y_\alpha} t_\alpha^b$. If all but the first one are irrelevant fields (that is: $y_\alpha < 0,\ \forall \alpha > 1$) it is useful to write the first coupling as

$$\beta_1^b - \beta_1^\star = s^{by_1} t_1 e_1^1 + \sum_{\alpha>1} s^{by_\alpha} t_\alpha e_1^\alpha \ . \tag{19}$$

For $\alpha > 1$, $y_1 - y_\alpha$ is typically near 2 (in the 3D XY model $y_1 \approx 1.5$, $y_\alpha \lesssim -1$) so that the second term in the RHS of (19) is negligible after some blocking steps.

If we restrict ourselves to a two parameter space $(\beta_1, \beta_2)$, the critical surface is approximately shown in figure 1. Starting on any point of the $S_\infty$ line, at each RG step the couplings move along it towards the fixed point corresponding to the particular RGT.

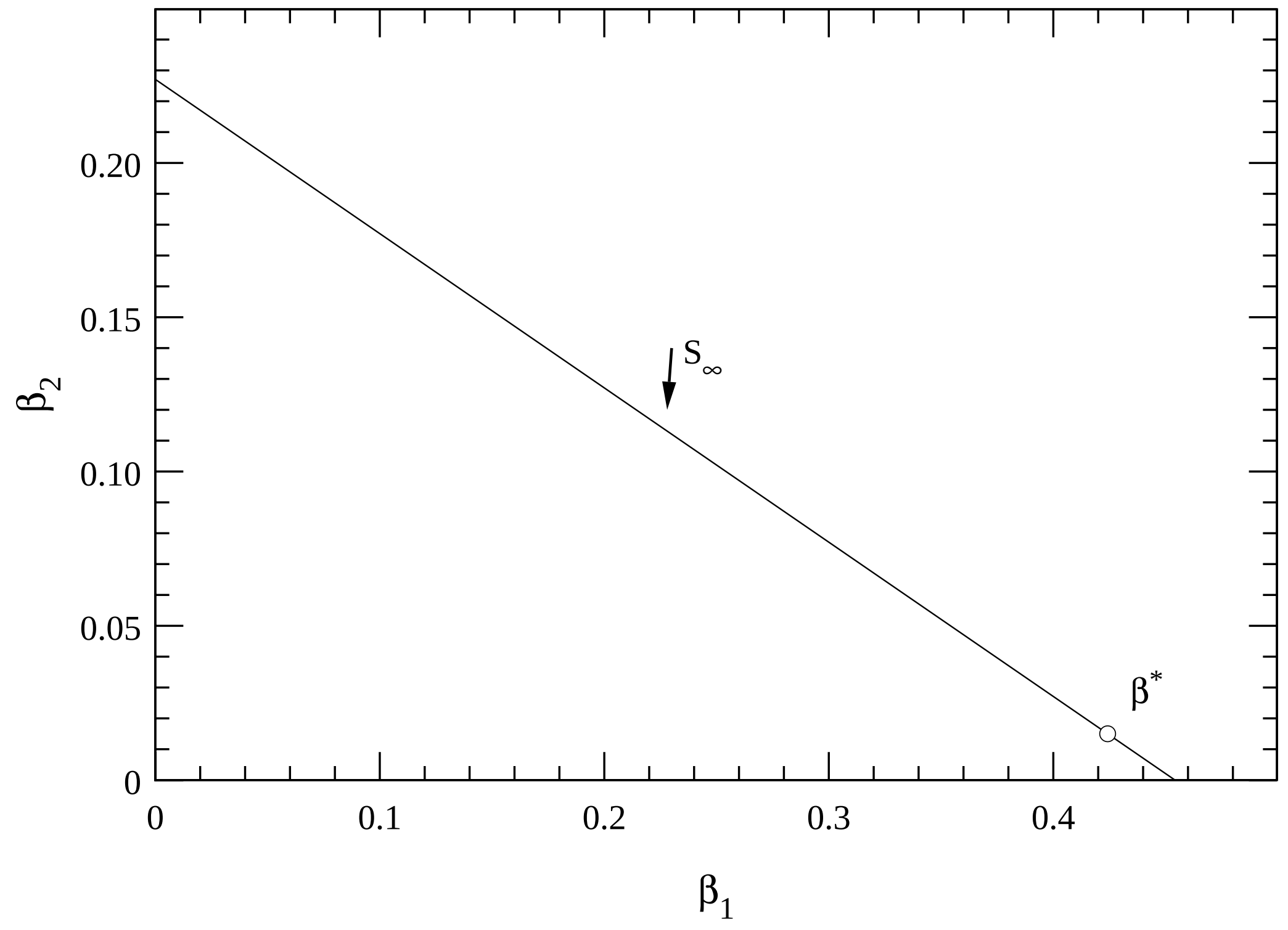

FIG. 1. Approximate representation of the critical surface of the $d = 3$ XY model in a two dimensional coupling parameter space. The fixed point lies in some point on the surface.

Close to $S_\infty$ but out of it, the relevant field is small and in the first RG steps its position will have small modifications, but as the field grows the point will rapidly move away from the critical line. The irrelevant field, in turn, will decrease. In this way we will follow asymptotically a line in the direction of the relevant eigenvector of the matrix $T$, with eigenvalue related with the critical exponent $\nu$. Drawing this flux it is readily seen where are located the fixed points of the transformation. In figure 2 we represent the flux obtained in four RG steps starting from different places close to the critical point. The points at $\beta_2 = 0$ represent the starting points, that is: points on the original lattice, including the simulation point itself as well as some neighbor points computed using the Spectral Density Method (SDM) [9]. After a RGT we obtain a lattice of side $L/2$ where applying the SDE we compute $\beta_1^1$ and $\beta_2^1$ . These points are linearly joined to the previous ones in figure 2 and the process is repeated for the following RGT.

In the first two steps we see, in figure 2, that the flux follows, with small corrections, the critical line towards the fixed point. It is clearly seen to be located between the second and third step, where the trajectories slightly start to separate from the critical line. In the fourth step they are rapidly moving away. A similar behavior for other RGT will be represented in figure 4.

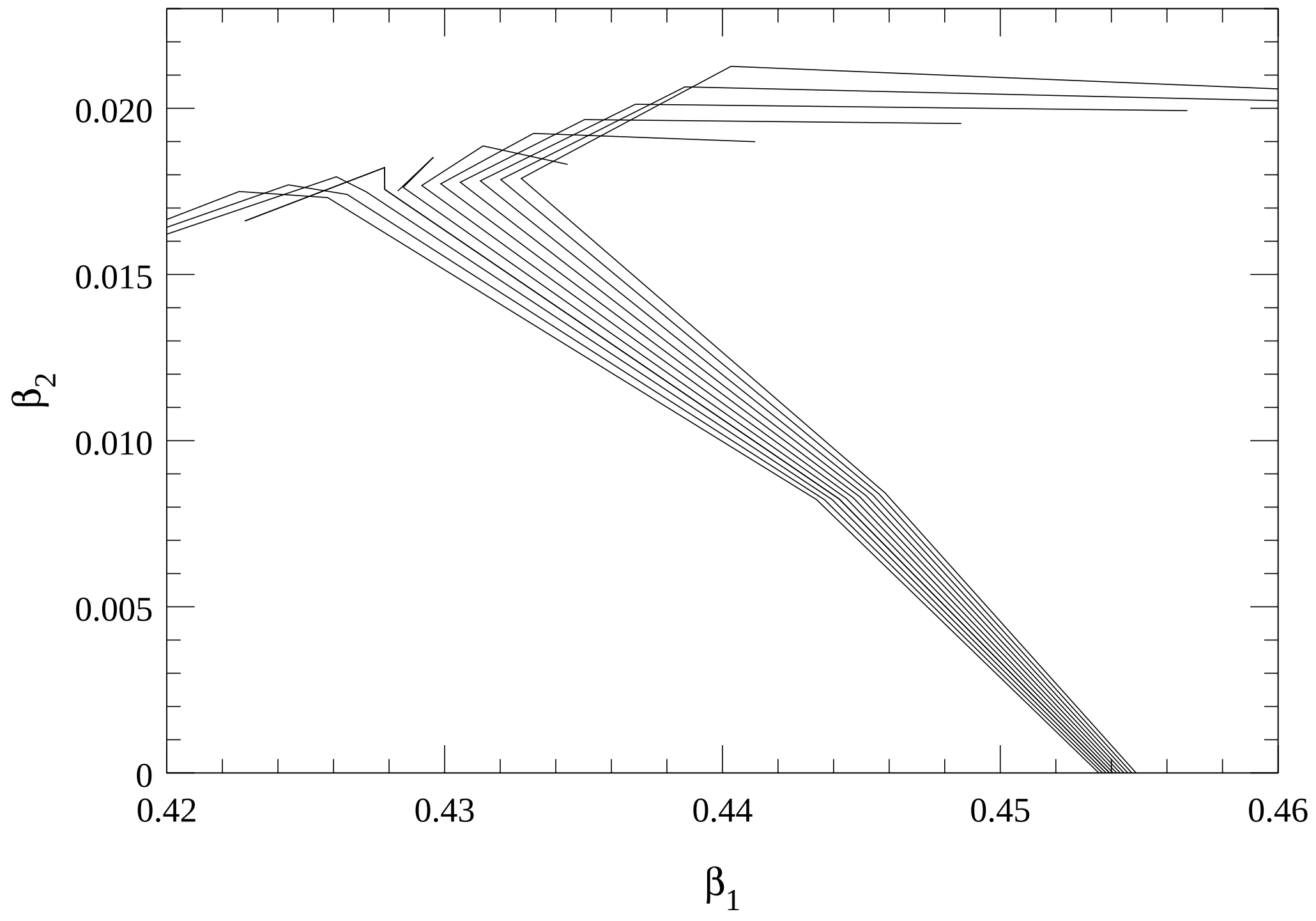

FIG. 2. Flux in the $(\beta_1, \beta_2)$ plane for starting points in a neighborhood of the critical (single coupling) point, in a $64^3$ lattice. For clarity we only plot the data until the next-to-last blocking level.

Our aim is to reduce the distance from the simulation point $\boldsymbol{\beta}_c = (\beta_c, 0, \ldots)$, to the fixed point $\boldsymbol{\beta}^* = (\beta_1^*, \beta_2^*, \beta_3^*, \ldots)$. As we have only two parameters to tune, we cannot vanish completely all $\beta_\alpha$, with $\alpha > 1$. It is a necessary condition for the proposed method to work that the fixed point may approach $(\beta_c, 0, 0, \ldots)$ with an appropriate selection of $n_s$ and $\epsilon$.

Numerically the complexity grows very fast with the number of couplings involved. First we will suppose that for all $\alpha > 1$ the couplings $\beta_\alpha^b$ are equal to zero, the stability of $\beta_1^b$ as a function of $b$ computed with the S-D equation (14) will give us an *a posteriori* confirmation of the reliability of that hypothesis. Second, we will consider that after the first RGT iteration only two non vanishing couplings $(\beta_1^b, \beta_2^b)$ exist. The absolute value of $\beta_2$ will be an estimation of the distance from the critical point to the simulation point.

We will present numerical results in the $d = 3$ XY model with lattice sizes ranging from $L = 8$ up to $L = 64$. We have mainly used the Wolff's Single Cluster algorithm [10]. We have measured in 100, 50, 40 and 10 thousands of configurations in $L$=8, 16, 32 and 64 respectively. Successive configurations are separated by a mean of 200 single cluster spin updates. We store every measure in order to compute the derivatives and to use the SDM. We have used jack-knife for error estimations.

### A. One coupling calculation

Let us make the hypothesis that $\boldsymbol{\beta}^* = (\beta_c, 0, 0, \ldots)$, and therefore let us use expression (14) to compute $\beta_1^b$. In the original lattice this expression also makes sense, and therefore we must obtain the same value for $\beta_1^0$, that is to say we should have $\beta_c = \beta_1^0$. If we use the SDM to move in the $\beta_1$ direction in a neighborhood of $\beta_c$, and plot $\beta_1^b(\beta_1)$, the point of matching of all couplings corresponds to the fixed point at this level of approximation. If we do not find the matching in a single point for all levels of $b$, this means that the fixed point has not been reached for this value of $(n_s, \epsilon)$. In this case, the fixed point is far from the simulation point and higher order couplings are not negligible.

In figure 3 we show the evolution of the blocked coupling as a function of the simulation coupling, obtained with the SDM, for several choices of $(n_s, \epsilon)$. The data have been taken in a $16^3$ lattice. We remark that the function $\beta_1^0$ obtained as a function of $\beta_1$ is effectively the identity function. The different parameters choices give different fixed points, but only if they are not far from the simulation point the truncated SDE will be accurate. For $n_s = 1$ we cannot obtain a good behavior for any value of $\epsilon$ (we plot in figure 3 the results with the *best* value). The results with a standard majority rule (summing the fields over $2^3$ cubes and normalizing the results) are of similar quality than for $n_s = 1$ (see figure 4 below). For $n_s = 2$ we plot the data obtained with two close values of $\epsilon$ to show the dependence. On the other hand, it may be also seen in figure 3 that increasing the number of smearing steps (see the results with $n_s = 4$) does not improve significantly the quality of the crossing, making useless the computational overload.

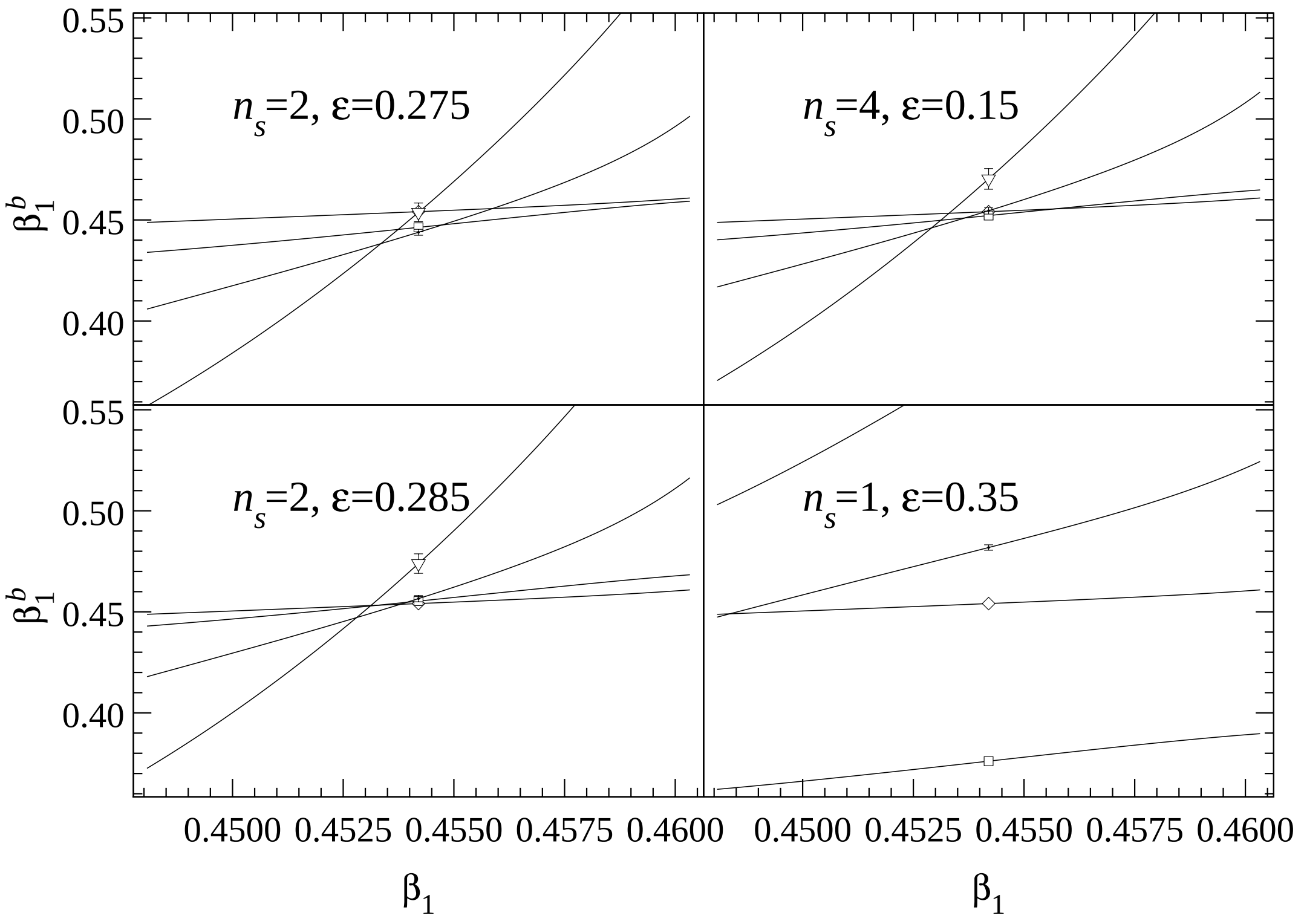

FIG. 3. $\beta_1^b$ computed with the Schwinger-Dyson equation (14) as a function of the simulation parameter $\beta_1$. The different lines correspond to different blocking level $b$. The slope grows with $b$. We show the results for several values of the RG parameters $(n_s, \epsilon)$. All the numerical data have been obtained from 6000 configurations of a $16^3$ lattice at the critical point.

### B. Two couplings calculation

After the renormalization transformation, we expect that there will be a set of renormalized couplings with non negligible values. To learn about the behavior of $\beta_i$ with $i > 1$ we will consider now just two nonzero couplings: $\beta_1^b, \beta_2^b$ corresponding respectively to the first and the second neighbors, that usually give the more important contributions. Now it is possible to draw the flux in a two dimensional parameter space.

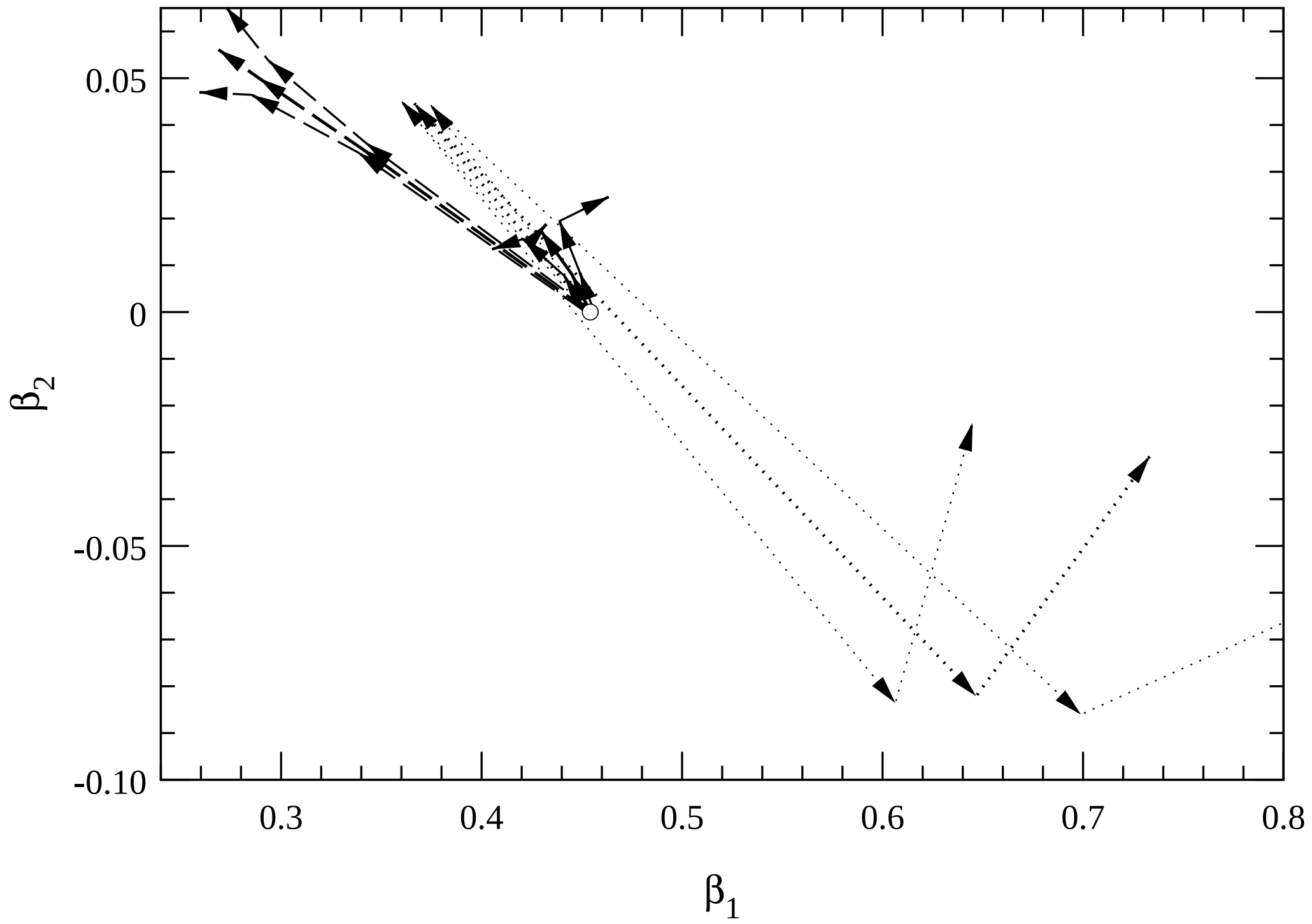

FIG. 4. Two dimensional flux for several RG transformations in a $32^3$ lattice. The solid lines correspond to the usual ($n_s = 2, \epsilon = 0.285$) selection, the dashed lines to ($n_s = 2, \epsilon = 0.2$) and the dotted line to a *majority rule* transformation without smearing.

In figure 4 we plot the two dimensional flux for some smearing transformations in a $32^3$ lattice. The solid lines correspond to the values ($n_s = 2, \epsilon = 0.285$). We have plotted the trajectories corresponding to starting points (0.4522,0), (0.4542,0) (central line for each set of trajectories) and (0.4562,0). To show the importance of the tuning of the $\epsilon$ parameter (in moving the fixed point), we also plot (dashed line) the trajectory with ($n_s = 2, \epsilon = 0.2$). The fixed point is one order of magnitude further. The situation is even worse when applying a simple majority rule, without smearing, (dotted lines), with a change in the direction of the flux. The numerical results for ($n_s = 1, \epsilon = 0.35$), not presented in figure 4, are again very similar to those from the majority rule.

The RGT performed in the following paragraph and sections will always correspond to the choice ($n_s = 2, \epsilon = 0.285$).

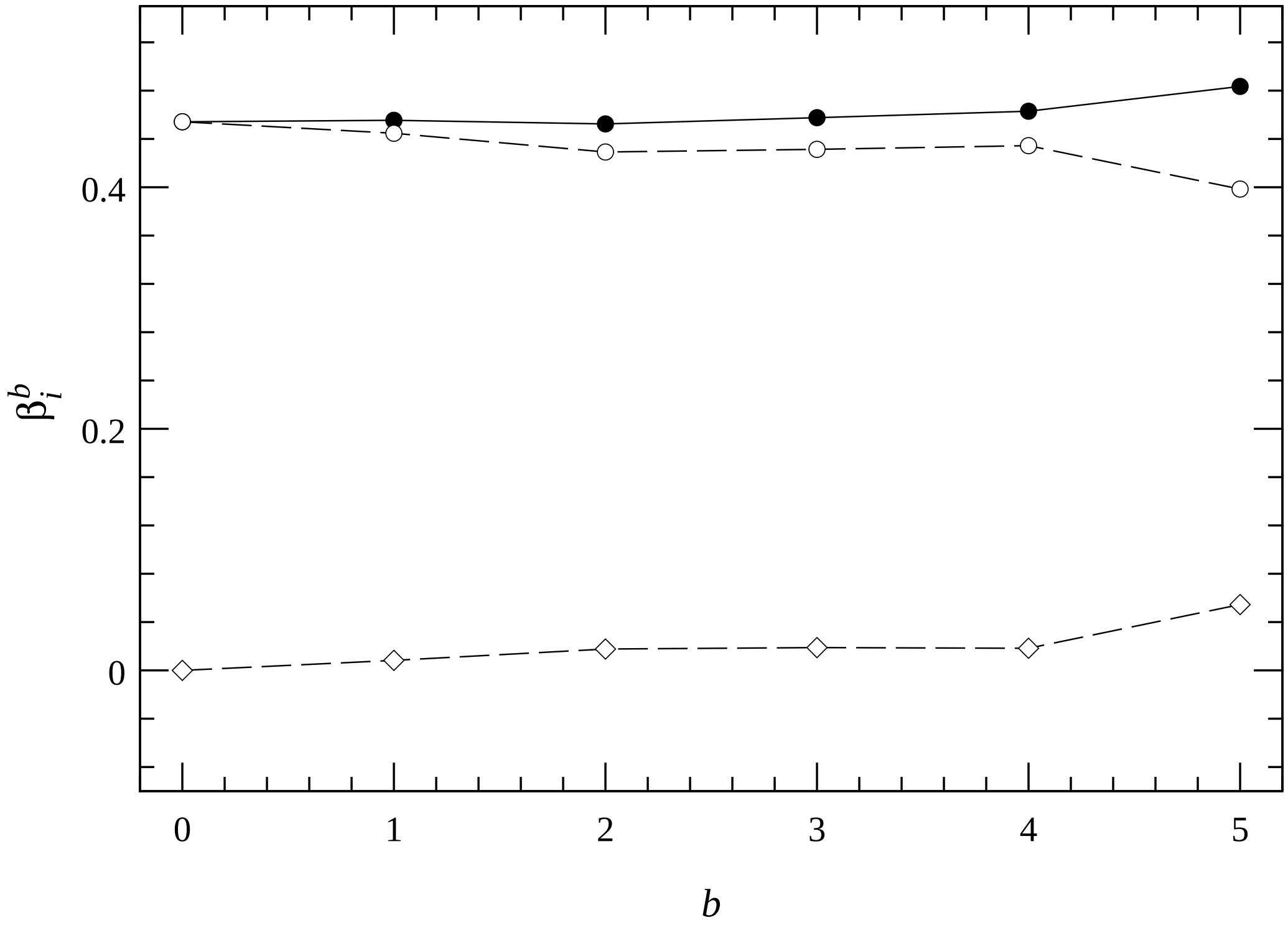

FIG. 5. $\beta_1$ (circles) and $\beta_2$ (diamonds) parameters as a function of the blocking level $b$ in a $64^3$ lattice. The solid line corresponds to the single coupling computation (equation (14)) and the dashed ones to the two coupling calculation (equation (16)). The error bars are smaller than the symbol sizes.

In figure 5 we show the evolution of the values of the couplings $\beta_1^b, \beta_2^b$ using the SDE in a $64^3$ lattice in the approximations of a single coupling ($\beta_1 \neq 0,\ \beta_i = 0,\ \forall i > 1$) and two couplings ($\beta_{1,2} \neq 0,\ \beta_i = 0,\ \forall i > 2$). In both cases we obtain a stable value after 2 transformations. The matching of the couplings for the second approximation is found at the second level of blocking, that is, we have reached the fixed point. On the other hand, the small variation of $\beta_1^b$ when including a second coupling in the SDE (about a 5%) shows the consistence of the single coupling approximation.

Due to finite size effects (see below) the latter transformation suffers from a large deviation. The result shows that with our selection of the smearing parameters we have $(\beta_1^\star, \beta_2^\star) \approx (0.43, 0.02)$. As we have started from the point $(\beta_1^\star, \beta_2^\star) = (0.4542, 0)$ the motion has been very small (the distance moved is similar to that schematically depicted in figure 1).

One may be tempted to tune $\epsilon$ in order to obtain $\beta_2^\star = 0$. But there, the SDE with $\beta_i = 0,\ \forall i > 1$ are also valid and will produce a deviation between $\beta_1$ and $\beta_1^\star = 0.4542$ that will be larger than the one obtained with $\epsilon = 0.285$. Remember that this value was selected to minimize the distance from the fixed point $\beta_1^\star$ in the one coupling calculation. We expect that the chosen value of $\epsilon$ will make small the higher order couplings at the fixed point.

### C. Systematic Errors

A first source of errors are the truncation effects that occur when the calculation is restricted to a single coupling space. Computations with more couplings may be useful to obtain higher precision results, and in particular may be efficient in a model as simple as the one we are considering here. However we are no strictly interested in reducing the truncation effects but in monitorizing them, for that reason we introduced the two coupling calculation. Notice that our main scope is to check the quality of the results when neglecting higher order contributions in order to know what can be expected when applying the method to more complex models. In particular those with interacting scalar and gauge fields, where the starting point is an action with several couplings, so that computing new renormalized higher order couplings may become a very complex task.

Another source of systematic errors is the possible nonlinearity of the RGT in the first RG steps if the starting point is not close to the fixed point. This effect can be reduced by discarding the measures at the first iterations.

Unfortunately the last RG steps may be also useless due to finite size effects. Let us consider for example the mean value of the energy operator. When the correlation length $\xi$ is near $L$, if we assume a correlation function of the type $G(r) = Ae^{(-r/\xi)}/r$, the contribution of a path that wraps around the lattice is of the order of $Ae^{-1}/L$, which is not negligible compared with the direct $G(1) \approx A$. This produces a growing of the value of energy-like observables ($O$) at the critical point that makes the crossing between $O^b(\beta)$ functions to shift to lower values of $\beta$. In other cases, the lattice size puts harder constraints on the observables, as for example happens for the mean value $\langle \cos(\theta_{n+\mu} - \theta_{n-\mu}) \rangle$ that becomes exactly 1 when $N_b = 2$.

However, when using equations (14) and (16) to compute the blocked couplings, the finite size effects are happily reduced giving reasonable values even at $N_b = 2$ when some operators involved in the computation of the couplings are completely saturated.

A quantitative estimation of finite size effects must be done comparing several lattices and blocking levels.

In the next two sections we will give some results regarding the computation of the exponent $\nu$, showing that all the systematic errors can be kept under the 3% level.

## V. THERMAL EXPONENT FROM THE RG FLUX

After the determination of the system flux diagram, one usually is interested in obtaining the critical exponents. We will now consider several methods to obtain the exponent $\nu$ studying the flux.

### A. Derivatives of the Renormalized Couplings

We can compute $\nu$ using the equation (19). However a direct use of (19) performing simulations near (but not *on* the critical surface ($t_1 = 0$)) is not convenient since the first term in the RHS of (19) grows very fast putting the renormalized coupling far from the critical region after a few iterations, and consequently loosing sense the linear approximation.

Alternatively, we can measure with a simple simulation at the single coupling critical point $(\beta_1^c, 0, 0, \ldots)$, the derivative of $\beta_1^b$ (14) with respect to $\beta_1$, just by computing the derivatives of the observables as the connected correlations with the intensive energy,

$$\frac{\partial \left\langle O^b \right\rangle}{\partial \beta_1} = L^d \left( \left\langle O^b E_1 \right\rangle - \left\langle O^b \right\rangle \left\langle E_1 \right\rangle \right) , \tag{20}$$

and with the SDM move in a region around this point.

From (19) we obtain

$$\frac{\partial \beta_1^b}{\partial \beta_1} = s^{b y_1} e_1^1 (D^{-1})_{1,1} + \sum_{\alpha > 1} s^{b y_\alpha} e_1^\alpha (D^{-1})_{1,\alpha} \ , \tag{21}$$

where $D_{ij} = e_j^i$. Notice that equation (21) is independent of the values of $t_\alpha$, with the restriction that they must be small enough to make valid the linear approximation.

If the second term in the RHS of eq. (21) is negligible (namely for $b$ large enough) we can write

$$\log \frac{\partial \beta_1^b}{\partial \beta_1} \approx b y_1 \log(s) + \log e_1^1 (D^{-1})_{1,1} \ . \tag{22}$$

Technically it is not possible to compute the $\beta_1^b$ just by measuring a reduced set of observables. In fact, the value obtained from equation (14) corresponds to the hypothesis of vanishing of the rest of the couplings. Let us now consider the possible bias introduced with this approximation in the computation of $\nu$. For a system at $\boldsymbol{\beta}$, using the SDE, we compute an approximation to the first coupling from the mean value of some simple observables that we will call $\beta_1^{SD}$. So we can write for any blocking step

$$\beta_1^{SD} = f(\beta_1, \beta_2, \ldots). \tag{23}$$

If $\beta_i$ are small $\forall i > 1$, we can linearly expand $f$ obtaining

$$\beta_1^{SD} \approx \beta_1 + \sum_{i>1} C_i \beta_i \tag{24}$$

where we have used the identity $\beta_1 = f(\beta_1, 0, 0, \ldots)$. In our results for the XY model the difference between $\beta_1^{b,SD}$ and $\beta_1^b$, $b > 1$ may be estimated as the difference between the one and two couplings calculation, that is about a 5%. Anyway, if we use (24) we obtain for the logarithm of the derivative of $\beta_1^{SD}$ an expression equivalent to (22) minus a variation of the independent term. The exponentially decreasing behavior of the rest of the terms in equation (21) remains with the only modification of multiplicative factors. In conclusion, the lack of a simple method for computing the couplings $\beta_1^b$ is not expected to be a source of bias.

Another effect that we could consider is the nonlinearity of the RGT. We expect to find this problem if the fixed point is far from the simulated critical point. We have confidence that the systematic error from this source is small since our transformation has the fixed point very near to the simulation one. However as we will see, an error in the 3% level cannot be excluded.

There is a simple method to learn about the importance of this effect, that is to compute the derivative of $\beta_1^b$ with respect to $\beta_1^{b'}$, with $b' < b$, from the derivatives

$$\frac{\partial \left\langle O^b \right\rangle}{\partial \beta_1^{b'}} = (L/2^{b'})^d \left( \left\langle O^b E_1^{b'} \right\rangle - \left\langle O^b \right\rangle \left\langle E_1^{b'} \right\rangle \right) \ . \tag{25}$$

Notice that if $b' > 0$ in equation (25) the computations are never done in the original variables, so that all measures are done nearer to the fixed point. On the other hand the derivative respect to $\beta_1^{b'}$ is computed directly from $E_1^{b'}$ (it depends on $\beta_1^b$ but not on $\beta_1^{b'}$).

## B. Numerical results

In figure 6 we show the evolution of $\log \frac{\partial \beta_1^b}{\partial \beta_1}$ as a function of the block number $b$ for the XY model. At $b = 0$ there is a deviation from the straight line (with slope $y_1 = 1/\nu$) due to the contribution of irrelevant fields. In the last blocking level the finite size effects are responsible of a new deviation.

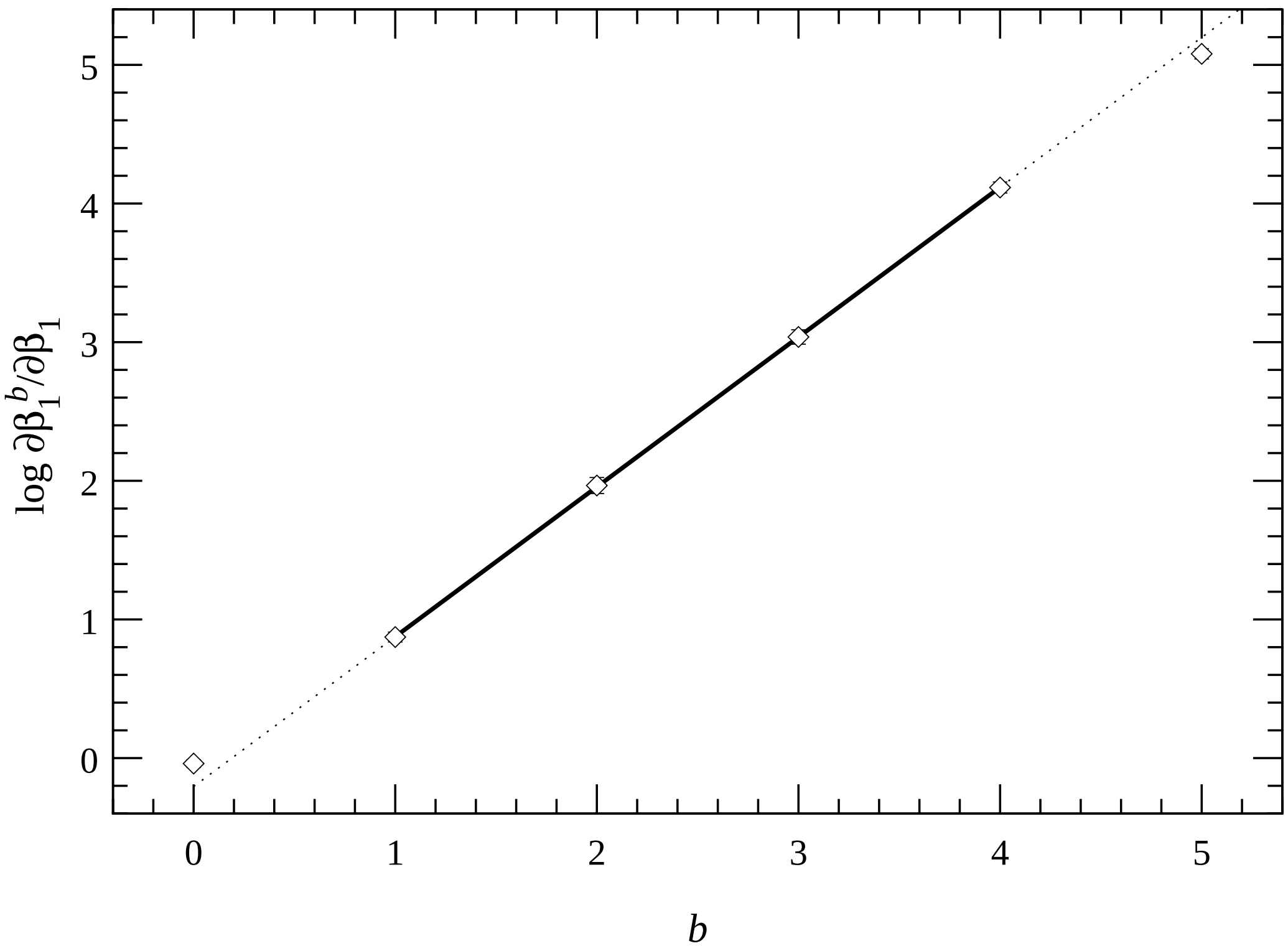

FIG. 6. $\log \frac{\partial \beta_1^b}{\partial \beta_1}$ computed from (20) as a function of the block level for the $64^3$ lattice.

In figure 7 we plot the results computing the ratio between derivatives of $\beta^{b'}$ and $\beta^b$ ($b' > b$) for all lattices (that is the slope joining any pair of points in figure 6 and equivalents). For $b = 1$ there are strong deviations for $b'$ small. For $b = 1, 2$ the statistical error is about a 2% while the systematic one is under the 3% level.

From linear fits to $\log \frac{\partial \beta_1^b}{\partial \beta_1}$ discarding the first and last points in the $L = 64$ and $L = 32$ lattices we obtain

$$\begin{aligned} \nu &= 0.638(10) \quad L = 64, \\ \nu &= 0.646(8) \quad\ \ L = 32, \end{aligned} \tag{26}$$

With a 1.5% of statistical error and about a 3% of systematic one (assuming $\nu \in [0.66, 0.67]$).

Using (25) to compute the derivatives with $b' = 1$ we obtain from the linear fitting of the points $b = 2, 3, 4$ in the $L = 64$ lattice,

$$\nu = 0.649(20) \tag{27}$$

with a 3% of statistical error and an unmeasurable systematic one, since it is compatible with the expected value.

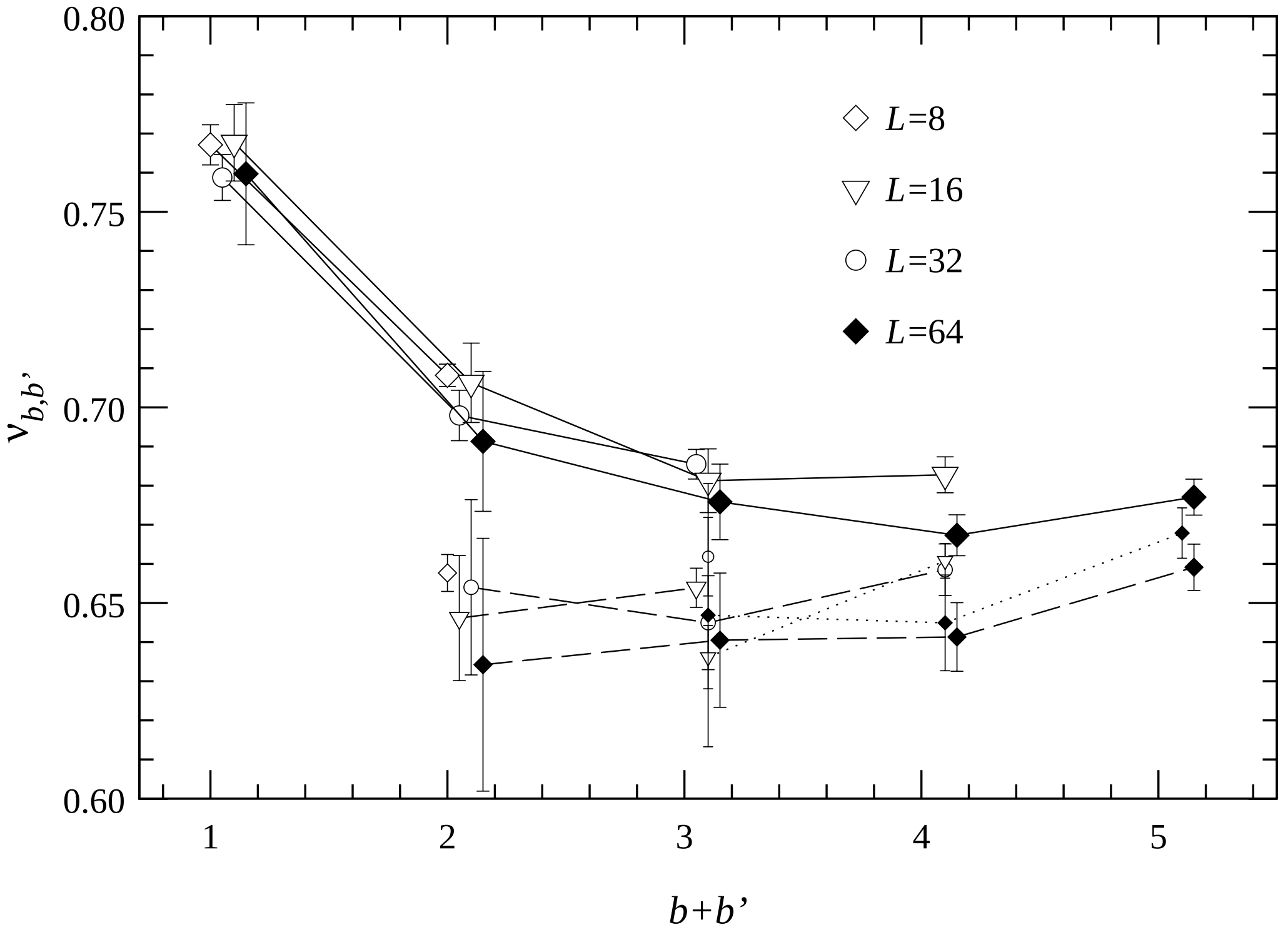

FIG. 7. Estimations of $\nu$ from the derivatives $\frac{\partial \beta_1^b}{\partial \beta_1}$ and $\frac{\partial \beta_1^{b'}}{\partial \beta_1}$ for all lattices. We only plot the points with $b = 0$ (larger symbols joined with solid lines), $b = 1$ (intermediate symbols and dashed lines) and $b = 2$ (small symbols and dotted lines).

Finally let us comment that another source of systematic error is the finite size effect over the critical point. Until now we have presented the results for $\nu$ obtained in the simulation point ($\beta_c = 0.4542$). In addition to a shift of the apparent critical point, the latter point itself is not well defined. From different definitions (namely the maximum of the derivative, the crossing point between couplings at different levels, etc.) we observe variations of the value of $\nu$ on the 1-2% level. For example, computing the derivative at the point where $\beta_1^b$ and $\beta_1$ match we obtain for the $L = 64$ lattice:

$$\nu = 0.654(11) \tag{28}$$

with a 2% of statistical error and a systematic one under the 2% level.

## VI. FINITE SIZE RENORMALIZATION GROUP

Another method to compute the exponent $\nu$ is to combine our RGT with a Finite Size Scaling (FSS) analysis [11]. Up to now, in order to compute fixed points, critical exponents, etc. we have been looking for matching, renormalized couplings..., always starting in a fixed lattice $L$, and blocking to $L/2, \ldots, 2$.

It is possible to carry out a very different study: taking data from two different lattices $L_1, L_2$ where RGT transformations are performed. After some steps, all irrelevant fields will be negligible. Comparing the results obtained from two lattices of original sizes $L_1$ and $L_2$, when different RG steps are taken, in order to end with the same final lattice, we can use the FSS techniques to obtain the critical exponent $\nu$. Applications of this method appeared in references [12–14].

In those works however, a single RG transformation reduces a $L^d$ lattice to (usually) a $2^d$ one. In principle, our method may be generalized to finite size blocks ($L/2$ length for instance) just by taking $n_s$ large enough, in order to let the system propagate the relevant information to all the block, avoiding a relevant lost of it after decimation. However, this would make this procedure too much time consuming.

The FSS ansatz affirms that in a finite system of length $L$ near the critical point, any dimensionless observable is a smooth function of $\xi/L$. In terms of the coupling we can write

$$\langle O \rangle_{L,\beta} = f(L^{1/\nu}(\beta - \beta_c)) \tag{29}$$

this means that the derivative at $\beta = \beta_c$ is just proportional to a power of the lattice size. Using data from lattice of sizes $L_1$ and $L_2$ we obtain (taking for simplicity equal final sizes)

$$\frac{1}{\nu} = \frac{\log \left( \frac{d}{d\beta} \langle O \rangle_{L_1,\beta} / \frac{d}{d\beta} \langle O \rangle_{L_2,\beta} \right) \Big|_{\beta=\beta_c}}{\log \frac{L_1}{L_2}} \ . \tag{30}$$

The procedure is then the following: we consider a $L_1$ lattice, and we block it up to a $L_f$ size. Now we start with a $L_2$ lattice and block it up to the same $L_f$ value. By using the observables computed in the $L_f$ lattices on the previous expression we obtain $\nu$.

The great advantage of the FSS method is that the finite size effects are no longer a source of systematic error that we need to fight against but the quantity we want to look at. For this reason we expect to obtain the more accurate values of $\nu$ in the maximum blocking level ($2^d$ lattice).

The FSS applies for all operators in the lattice. We can consider the previous renormalized couplings that are functions of the neighbor correlators, but also the latter operators themselves.

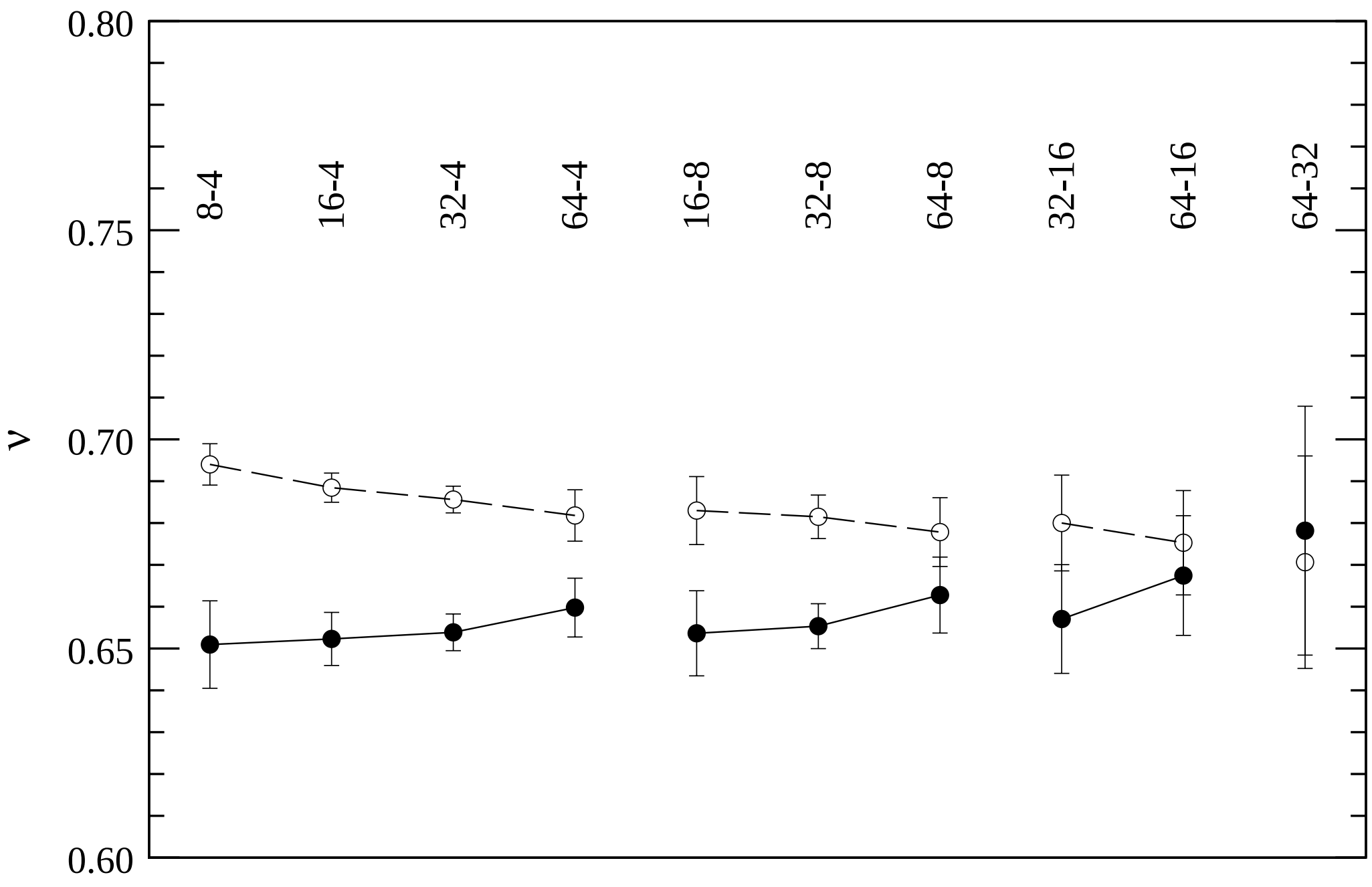

FIG. 8. Values for the critical exponent $\nu$ obtained from a Finite Size Scaling analysis of the couplings at the last ($L_f = 2$) level of blocking for the pairs $L_1 - L_2$ displayed. The white points have been obtained from first neighbor energy operator and the filled ones from the corresponding coupling. We have also included data of a very small lattice ($L = 4$) as a control.

In figure 8 we plot the results obtained using the energy (next neighbor correlation), as well as the value of the coupling obtained from (14). We observe a clear systematic error for the small lattices with opposite sign for the energy and the coupling. At sizes larger than $16^3$ the systematic error is under the statistical one with a total error under the 3% level.

## VII. CONCLUSIONS

We have proposed a Real Space Monte Carlo Renormalization Group transformation whose main features are

- It is easy to implement for many systems, even complex ones, since we only need to define a relaxation procedure.
- With a 3% of precision level we can neglect the truncation effects, at least in the 3-$d$ XY model.
- The code for the transformation is very easy to implement, since the more time-consuming part can be done with a slight modification of what one usually does in a local Monte Carlo iteration.
- The adaptation to parallel computers is straightforward since most of the needed operations are local.

As a next step we want to test the method in a gauge theory. Unfortunately the more simple gauge theories, with continuous groups, have no critical (second order) points at finite values of the coupling. We project to study gauge fields coupled to matter (namely the U(1)-Higgs model).

However, we have performed some calculation in the four dimensional U(1) model at the (first order) Confinement-Coulomb transition. The results show a good behavior, regarding the stability of the coupling, after an appropriate tune of the parameters of the transformation.

We acknowledge CICyT for partial financial support with the projects AEN93-0604, AEN93-0776 and AEN94-218. One of us (JJRL) is granted by MEC (Spain).